\documentclass{emulateapj}
\usepackage{lscape}
\input{epsf.sty}
\def\solar{\ifmmode_{\mathord\odot}\else$_{\mathord\odot}$\fi} % sun
\def\kms{km\thinspace s$^{-1}$}     % km s-1
\def\deg{\ifmmode^\circ\else$^\circ$\fi}  %degree sign
\def\arcs{\ifmmode {'' }\else $'' $\fi}  % arc se
\def\arcm{\ifmmode {' }\else $' $\fi}    % arc min

\def\msolar{M$_\odot$}
\def\hi{\ion{H}{1}}
\newcommand{\gapprox}{\ifmmode \buildrel > \over {_\sim} \else $\buildrel >\over {_\sim}$\fi}
\newcommand{\lapprox}{\ifmmode \buildrel < \over {_\sim} \else $\buildrel <\over {_\sim}$\fi}

\begin{document}

\title{The Nearby Damped Lyman-$\alpha$ Absorber SBS 1543+593: A Large \hi\
Envelope in a Gas-rich Galaxy Group}

\author{Jessica L. Rosenberg\footnote{National Science Foundation Astronomy and 
Astrophysics Postdoctoral Fellow and visiting astronomer, Kitt Peak National 
Observatory, National Optical Astronomy Observatory, which is operated by the 
Association of Universities for Research in Astronomy, Inc. (AURA) under 
cooperative agreement with the National Science Foundation.}} 
\affil{Harvard-Smithsonian Center for Astrophysics, 60 Garden Street,
Cambridge, MA 02138}
\author{David V. Bowen} 
\affil{Princeton University, Peyton Hall, Ivy Lane, Princeton, NJ 08544}
\author{Todd M. Tripp}
\affil{Department of Astronomy, University of Massachusetts, Amherst, MA 01003}
\author{Elias Brinks}
\affil{Centre for Astrophysics Research, University of Hertfordshire, College
Lane, Hatfield AL10 9AB, U. K.}

\begin{abstract}
We present a Very Large Array\footnote{The Very Large Array is part of the National 
Radio Astronomy Observatory which is a facility of the National Science 
Foundation operated under cooperative agreement by Associated Universities, Inc.} 
(VLA) \hi\ 21cm map and optical observations of the region around one of the 
nearest damped Lyman-$\alpha$ absorber beyond the local group, SBS 1543+593. 
Two previously uncataloged galaxies have been discovered and a redshift has been
determined for a third. All three of these galaxies are at the redshift of SBS 
1543+593 and are \lapprox 185 kpc from the damped Lyman-$\alpha$ absorber. We 
discuss the \hi\ and optical properties of SBS 1543+593 and its newly identified 
neighbors. Both SBS 1543+593 and Dwarf 1 have baryonic components that are
dominated by neutral gas -- unusual for damped Lyman-$\alpha$ absorbers for
which only $\sim$5\% of the \hi\ cross-section originates in 
such strongly gas-dominated systems. What remains unknown is whether low mass
gas-rich groups are common surrounding gas-rich galaxies in the local universe
and whether the low star-formation rate in these systems is
indicative of a young system or a stable, slowly evolving system. We discuss
these evolutionary scenarios and future prospects for answering these questions.
\end{abstract}


\section{Introduction}

Damped Lyman-$\alpha$ (DLA) absorbers are clouds of high column density neutral
hydrogen gas seen in absorption in the spectrum of a bright background source,
usually a QSO. These are the highest column density absorption systems and are
generally found in galaxies where the gas is predominantly neutral (for a detailed 
description and review of DLA studies see \citealp{wolfe2005}). These absorption
line systems are particularly useful for tracing the densest neutral hydrogen
gas, and thereby tracing gas-rich galaxies, throughout the universe.

DLA observations are unique in that they allow us to detect high redshift 
galaxies in a manner that is almost entirely independent of galaxy luminosity --
the requirement to detect the DLA is a bright background QSO and a high enough
column density of \hi\ gas in the foreground galaxy. In this way a large number
of gas-rich galaxies and/or protogalaxies have been identified in the distant
universe. In order to determine the properties of these systems, \citet{wolfe1998} 
and \citet{prochaska1997} have used the kinematics of DLA absorption lines to
argue that DLAs originate in massive spiral galaxies. However, there is a
growing body of evidence that these systems span the range of gas-rich galaxy 
morphologies (e.g., \citealp*{rosenberg2003,bowen2001a,bouche2001,steidel1994, 
lanzetta1997,lebrun1997, pettini2000, turnshek2001, colbert2002, kulkarni2005}).
DLAs are gas-rich galaxies some of which are massive, luminous, and easy to
detect, but many others are compact, low luminosity, or low surface brightness 
systems that are difficult to detect at large distances.

Because many of the galaxies associated with DLAs remain undetected, it can be 
difficult to make the connection between the dense neutral gas and the
properties of the galaxy within which it resides. For this reason, the small
number of DLAs for which the associated galaxy has been identified provide 
important information about the properties and environment of DLAs. In this
paper we present a discussion of the properties and environment of SBS 1543+593, 
one of these select systems for which the associated galaxy has been identified.

SBS 1543+593 is one of the nearest known DLAs outside of the local group (only
NGC 4203/Ton 1480 is closer, \citealp{miller1999}). A QSO was 
discovered at the position of this system as part of a QSO survey. Subsequently
it was realized that the QSO lay behind this nearby foreground galaxy 
\citep{reimers1998}. The galaxy is a low surface brightness galaxy (LSB;
$\mu$(R) = 22.4; \citealp{bowen2001a}) with an HI mass of 
$\sim1.3\times10^9$ M\solar\ \citep{bowen2001b, 
chengalur2002}. The QSO sightline has provided a probe of the interstellar medium
in the foreground galaxy as discussed in detail in \citet{bowen2005}.

The galactic environment in which a DLA resides can have a significant impact on
its properties and its evolution. However, the galactic environment of DLAs, 
particularly on small scales and in the local universe, remains largely unknown.
For one system at $z_{abs} = 3.39$, at least 4 galaxies have been detected
within 5 $h^{-1}$ Mpc and the \hi\ in the DLA is found to be highly turbulent
\citep{ellison2001}. At $z \sim 3$ there have also been a number of studies 
of the clustering of DLAs. Several of these studies failed to detect 
any clustering of Lyman-break galaxies (LBGs) with DLAs at $z \sim 3$ 
\citep{gawiser2001, bouche2004} because of poor statistics. However, other larger 
studies arrive at inconsistent results on the 
clustering of these populations -- \citet{cooke2006} find that the DLA - LBG cross 
correlation is comparable to the LBG autocorrelation implying that DLAs are
strongly clustered massive halos while \citet{adelberger2003} find that the DLA
- LBG clustering implies that the DLAs are less strongly clustered and therefore
reside in lower mass halos. 

Because of the importance of environment on the evolution of galaxies,
we use the data on SBS1543+593 not only to probe the properties of this nearby
DLA but also to probe its gaseous environment. Previous studies of low surface 
brightness galaxies like SBS 1543+593 have shown that DLAs are generally 
less clustered on scales of a few Mpc than their higher surface brightness
counterparts \citep{rosenbaum2004,mo1994}. Nevertheless, it is not unusual for LSBs 
to reside in small groups on scales of $\sim$ 0.5 Mpc \citep{bothun1993}.
The low redshift of SBS 1543+593 makes it possible to study the environment of
this low surface brightness DLA in detail. 

In this paper we use VLA 21 cm observations to map the neutral gas in and
around SBS 1543+593 and present $V$-band data for its three newly identified 
companions. We discuss the VLA and optical observations of SBS
1543+593 and the new companions in \S 2. In \S 3 we present the
results including a discussion of the relationship between the \hi\ and optical
distributions in SBS 1543+593 (\S 3.1) and in the newly identified dwarf
galaxies (\S 3.2). Most of the work that has been done on the environment of 
LSB galaxies has relied on the optical detection of their neighbors. \hi\ 
observations provide a means of tracing the gas in the region. In \S 3.3 we 
discuss the small scale \hi\ environment and large scale structure in this region. 
Finally, in \S 4 we discuss the impact of our findings.

We assume a value of H$_0 = 70$ \kms\ Mpc$^{-1}$ throughout this paper.

\section{Observations}
\subsection{21cm Data}

\begin{figure}[ht]
\plotone{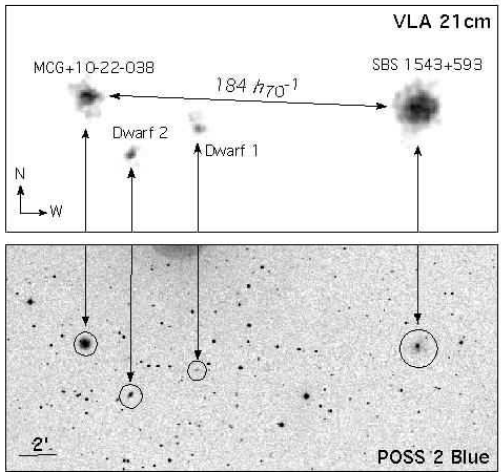}
\caption{A view of the field around SBS 1543+593 showing the relative positions
of the galaxies in this group at 21 cm from the VLA maps (upper panel) and in 
the POSS2 blue image taken from the STScI Digitized Sky Survey (lower panel).}
\label{fig:region}
\end{figure}

We use VLA C-array data of SBS 1543+593 to study the \hi\ distribution and 
environment of this nearby DLA. Within these data we identify SBS 1543+593, 
the low surface brightness galaxy observed as a DLA along the sight line to the 
QSO HS 1543+5921 ($z= 0.807$) and 3 previously unknown companions within
$\sim$185 kpc of the galaxy (see Figure \ref{fig:region} for a picture of the region). 

The VLA data were taken on August 13th and 14th, 
2001 and consist of 283 minutes of observations on the source
plus observations of 3C286 as a flux and bandpass calibrator and of 3C343 as a
complex gain calibrator. The observations were made in 2 IF mode with the bandpass 
centered at 1.4069 GHz (2971 \kms), close to the systemic velocity of the galaxy
measured from single dish observations by \citet{bowen2001b}. One IF covered a 
bandwidth of 1.5 MHz with a velocity resolution of 5.0 \kms channel$^{-1}$. The 
second IF covered 
a bandwidth of 3.0 MHz at a velocity resolution of 20.6 \kms channel$^{-1}$. 

The high and low velocity resolution data were both reduced using standard AIPS
data reduction techniques. The high velocity resolution data, which contain only 
SBS 1543+593 itself, were fourier transformed using the task IMAGR with ROBUST=0 
(a compromise between uniform weighting and natural weighting) which resulted in 
a final cube with a CLEAN beam of 15\arcs\ $\times$ 14\arcs. Moment maps were created 
from this cube using MOMNT to integrate over the channels which show line emission
and an equivalent single dish spectrum was created by running ISPEC for
the spatial region of the cube in which emission is detected. The lower resolution 
data were used for the observations of SBS 1543+593's neighbors. These data were 
fourier transformed using IMAGR with natural weighting and produced a final cube 
with a 19\arcs\ $\times 18$\arcs\ CLEAN beam. 

The newly detected galaxies were far from the field center and were, therefore,
significantly affected by primary beam attenuation. In
order to extract all of the signal while minimizing the spurious noise spikes,
the low velocity resolution data were analyzed as follows: (1) the
imaged data cube was convolved with a 30\arcs\ $\times$ 30\arcs\ beam that was
then clipped to keep only the data above 2-$\sigma$ (0.7 mJy), (2) all emission
that was not correlated over a few channels as identified by-eye with TVMOVIE, 
was clipped as noise since genuine emission should exist over several velocity 
channels, (3) the cube that resulted from steps 1 and 2 was used to select the 
regions of emission from the original unsmoothed data cube using BLANK, (4) the 
resulting data cube was then used to make moment maps (using XMOM) and 
equivalent single dish spectra (using ISPEC). Note that the data blanking for
the low velocity resolution cube means that pixels in which there is no
detectable \hi\ emission were blanked leaving signal only in the limited number of
pixels where significant galaxy emission was detected.

The rms noise in these data cubes is 0.35 mJy. 
For a 50 \kms\ wide source at the center of the field, a galaxy with 3.4
$\times 10^7$ M\solar\ of \hi\ would be a 5$\sigma$ detection at 40.7 Mpc (the
distance of SBS 1543+593). The companions detected near SBS 1543+593 
include MCG+10-22-038, a previously cataloged galaxy without a known redshift 
and two previously uncataloged systems, RBTB154542.8+591132 and 
RBTB154607.7+591013. For simplicity we will refer to the latter two galaxies as 
Dwarf 1 and Dwarf 2. The rms in the region of these detections falls to
$\sim$1.6 mJy corresponding to $\sim$1.6$\times 10^8$ M\solar\ so these newly 
identified dwarfs are near the detection limit at this distance from the center 
of the field. 
Figure \ref{fig:region} shows optical and \hi\ 21 cm maps of this region. The
lower panel is the POSS2 blue image of the region; the upper panel is the 
corresponding VLA \hi\ map. The three new galaxies form a tight group slightly 
offset from SBS 1543+593. No other galaxies were identified in the VLA data.

\begin{figure}[ht]
\plotone{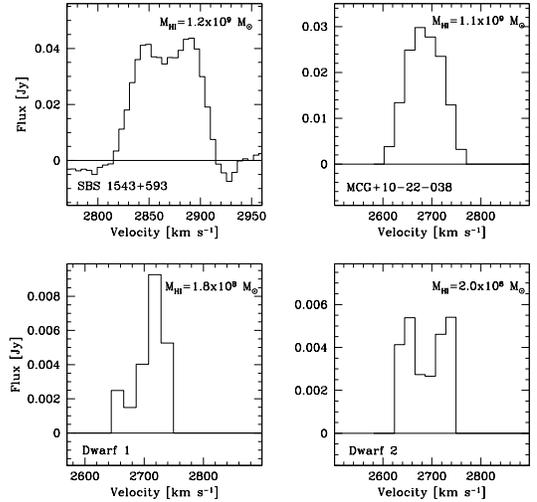}
\caption{Spectra for SBS 1543+593, MCG+10-22-038, Dwarf 1, and Dwarf 2. The 
velocity resolution for the SBS 1543+593 spectrum is twice as high as that of
the other galaxies. The velocity range of this galaxy was covered by the IF with 
a smaller bandwidth and higher velocity resolution. The other spectra were derived 
from the larger bandwidth, lower resolution data. The spectra of MCG+10-22-038, 
Dwarf 1, and Dwarf 2 do not show any baseline noise because of the clipping that 
was applied to the data (see \S2.1 for details).}
\label{fig:HIspec}
\end{figure}

The equivalent single dish spectra for the galaxies are shown in Figure 
\ref{fig:HIspec}. The spectrum of SBS 1543+593 was created from the higher 
velocity resolution smaller bandwidth IF data. The companion galaxy spectra 
were made from the larger bandwidth, lower velocity resolution data. The lack of
noise in the baseline of these lower resolution spectra is a result of the noise 
clipping described above. Figure 
\ref{fig:HIspec} shows that the companion galaxies all have similar recession
velocities which are offset from the velocity of SBS 1543+593 by $\sim$150 \kms. 
The velocities, velocity widths, and \hi\ masses for the 
galaxies are calculated from these spectra and the results are presented in
Table 1. We assume a distance of 40.7 Mpc for all of the sources based on the
recession velocity of SBS 1543+593.

The \hi\ sizes for these galaxies were all computed at the 1 M\solar\ pc$^{-2}$
level. The sizes were computed by fitting isophotal contours to the data using
the ELLIPSE routine which is part of the STSDAS package in IRAF. Except for
Dwarf 2 which is well fit by elliptical contours, we fit the galaxies with 
circular isophotes to determine the sizes. 

For SBS 1543+593 we measure an \hi\ mass of 1.2$\times 10^9$ M\solar, 
consistent with the values measured by \citet{chengalur2002} and
\citet{bowen2001b}. MCG+10-22-038 is the only previously known galaxy of the three 
neighbors to SBS 1543+593, but it did not have a cataloged redshift so its 
proximity to the DLA was not previously known. Therefore, MCG+10-22-038 has an
\hi\ mass of 1.1$\times 10^9$ \msolar while Dwarf 1 and Dwarf 
2 are lower mass systems with \hi\ masses of 1.8 $\times 10^8$ 
\msolar\ and 2.0$\times 10^8$\msolar\ respectively.

\subsection{Optical Data}

The optical data for SBS 1543+593 and its neighbors consist of $R$-band imaging
from the Apache Point 3.5m for all of the objects except for MCG+10-22-038 (the
only previously cataloged neighbor) and $V$-band imaging from the Kitt Peak 0.9m 
for all of SBS 1543+593's neighbors. The optical properties of SBS 1543+593 are 
discussed in detail in \citet{bowen2001b} and are included here for discussion
with respect to the properties of the neighbors. We also use the image of SBS
1543+593 taken with the Gemini Multi-Object Spectrograph r-G0303 (exposure time
1800 seconds) by \citet{schulte2005} for comparison with the HI data.

Standard calibration frames and a single $V$-band image of the region
surrounding MCG+10-22-038 were obtained at the Kitt Peak 0.9m in April 2002. 
Standard IRAF procedures were used to bias subtract, flat field, 
illumination correct and cosmic ray clean the image. For a complete description
of the flux calibration see \citet{stevenson2006}. In order to subtract the
background before photometry could be performed, a five pixel wide annulus was
placed around each galaxy beyond the edge of the emission from the object. The
median value from this annulus was used to subtract the background. Aperture 
photometry was then performed on each galaxy using ELLIPSE. For MCG+10-22-038 
and Dwarf 1 we used circular apertures while for Dwarf 2 we use an elliptical
aperture with an ellipticity of $0.52$ and a position angle of $-29.3$. 

The optical data that exist for these galaxies are R-band and V-band
measurements. However, because they tend to be blue systems, low surface
brightness galaxies are usually (in the literature) measured at B-band. In order
to compare with results from the literature we assume colors for the conversion 
of the central surface brightnesses and magnitudes to B-band. For the $B-V$
color of SBS 1543+593's neighbors we assume the median color for dwarf low surface 
brightness galaxies from the sample of \citet{vanzee1997}, $B-V = 0.49$. For the
$B-R$ color of SBS 1543+593 we assume the mean color, $B-R = 0.78$, for 
LSBs (not dwarfs) from \citet{deblok1995}. 

Optical diameters of the galaxies were determined in three different ways
to allow us to make comparisons with data from the literature. The measurements
of these diameters are included in Table 2: (1) The galaxy diameter within which 
we compute the total magnitude. The diameter of the aperture was determined by 
fitting apertures until the flux stopped growing, indicating that we 
had hit the background (which was well inside the annulus used to fit the 
background). For Dwarf 1 and Dwarf 2, this aperture method was applied to the 
$V$-band images and then the same aperture was used to derive the total $R$-band
flux (D$_{tot}$). (2) The diameter where the galaxy reaches an isophotal level 
of 25 mag arcsec$^{-2}$ (D$_{25}$). (3) The diameter measured at 6.4 times the
scale length of the galaxy 
(D$_{6.4h}$). As with the magnitudes, we have measured diameters in different
bands than the literature values. In order to convert from our
measurement of isophotal diameter to B-band isophotal diameter we use 
D$_{25}^R$/D$_{25}^B = 1.4$ from \citep{swaters2002} and, by extrapolation, that 
D$_{25}^V$/D$_{25}^B = 1.2$. The galaxy scale length is assumed
to be independent of wavelength in this optical range as was found by
\citet{swaters2002} for their late-type dwarf galaxy sample.

\begin{figure}[ht]
\plotone{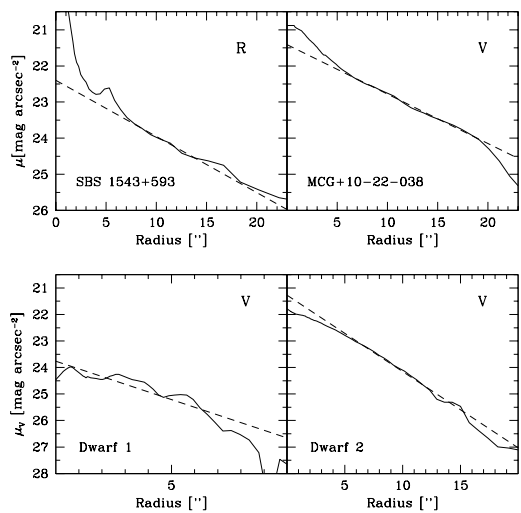}
\caption{The $R$-band surface brightness profile of SBS 1543+593 and the $V$-band 
surface brightness profiles for the three galaxy neighbors
to SBS 1543+593. The solid lines show the surface brightness profiles derived
from the images, the dashed lines show exponential fits to the data. In the
surface brightness profile of SBS 1543+593 the extremely steep rise in the
inner-most region is due to the QSO, the bump in the profile at $\sim$5\arcs\
is a star.} 
\label{fig:profs}
\end{figure}
 
Figure \ref{fig:profs} shows the $R$-band surface brightness profile for SBS
1543+593 and the $V$-band surface brightness profiles for MCG+10-22-038, Dwarf
1, and Dwarf 2. Fits to the exponential region of the surface brightness
profiles are shown with dashed lines. The central surface brightnesses
(corrected to $B$-band) listed in Table 2 are the extrapolation of this fit to
the center of the galaxy. All of these galaxies have low central surface 
brightnesses. The surface brightness rises sharply towards the center of SBS 
1543+593 because of 
the combination of light from the nucleus and the background QSO. For SBS 1543+593
we have determined the B-band central surface brightness assuming (B-R) = 0.2 
\citep{schulte2004}. Dwarf 1 has a 
$B$-band central surface brightness (assuming B-V = 0.49) of $\mu_0$(B) $= 24.3$ 
mag arcsec$^{-2}$ and a flat surface brightness profile. MCG+10-22-038 has a
small R$^{1/4}$ bulge but a disk central surface brightness of $\mu_0$(B) $=
21.9$ mag arcsec$^{-2}$. Dwarf 2 is unusual in having a slight depression in the 
central surface brightness with the surface brightness only reaching $\mu_0$(B) 
$= 22.2$ mag arcsec$^{-2}$ while the extrapolated exponential disk surface 
brightness is $\mu_0$(B)$ = 21.8$ mag arcsec$^{-2}$.

\section{Results}

\subsection{The Relationship Between Optical and \hi\ Emission in SBS 1543+593}

\begin{figure}[ht]
\plotone{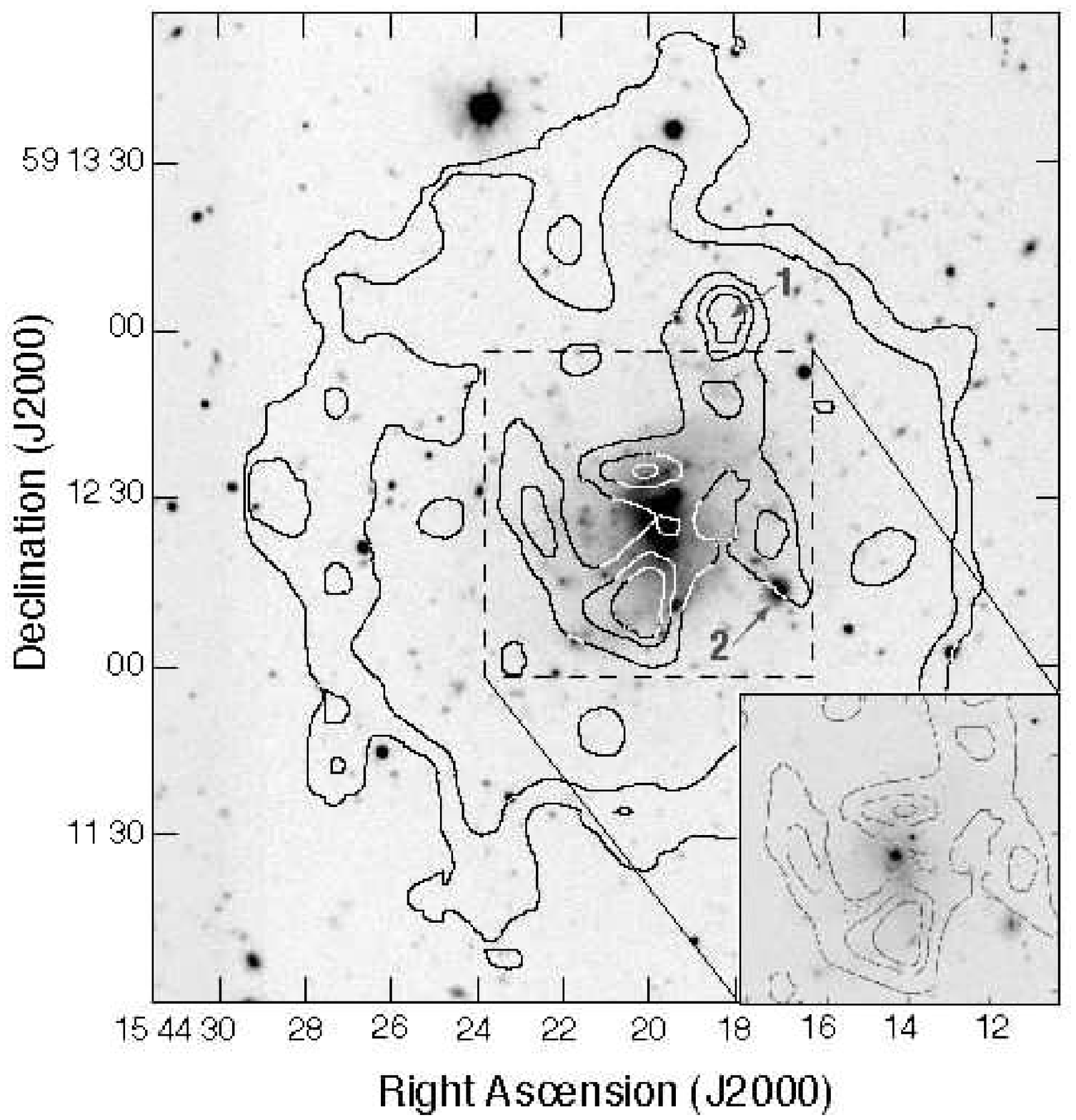}
\caption{The GMOS r-G0303 image of SBS 1543+593 from
\citet{schulte2005} overlaid with \hi\ contours from the VLA map. The lowest level 
contour represents 2$\times 10^{20}$ cm$^{-2}$, the column density above which the
gas would result in a damped Lyman-$\alpha$ absorber if it were in the
foreground of a QSO. The subsequent contours are 6, 16, 20, and 22 $\times 
10^{20}$ cm$^{-2}$. Region 1 is a high HI column density region that shows
little optical emission. Region 2 is a background spiral galaxy that is more
easily made out as a spiral galaxy in the inset STIS image.}
\label{fig:sbsmap}
\end{figure}

The relative distributions of the optical and \hi\ emission allow us to examine 
the relationship between stellar emission and gas density in a galaxy. Figure 
\ref{fig:sbsmap} shows the \hi\ distribution in SBS 1543+593 overlaid on the GMOS 
r-G0303 image of SBS 1543+593 from \citet{schulte2005}. The lowest level contour 
in the \hi\ map is at 2$\times 10^{20}$ cm$^{-2}$, the column density above which 
the gas would result in a damped Lyman-$\alpha$ absorber if it were in the 
foreground of a QSO. This contour represents a 4-$\sigma$ detection in a single 
channel of this map (the figure shows the integration of 20 channels). 

The \hi\ distribution shown in Figure \ref{fig:sbsmap} is consistent 
with the map of \citet{chengalur2002}. The highest column density \hi\ in the central 
region of the galaxy follows the optical light. Two
regions of particular interest have been labeled in Figure \ref{fig:sbsmap}.
Region 1 is outside of the main spiral structure, 
has high gas density, and is a region of low surface brightness stellar
emission. This emission comes from slightly farther out than the spiral arms
that are visible and appears to point in the opposite direction. The reason for
this disturbance in the spiral arm structure is unclear but could point to a
minor interaction in the past. The bright optical emission located near region 2 
is a background spiral galaxy (discernible in the inset STIS image of the
region). The most striking feature in the combination of the
optical and \hi\ images is that there is a large fraction of the \hi\ disk that
shows little evidence for stellar emission even in this deep GMOS image.
Significant asymmetries exist in the inner regions of the galaxy and more 
minor asymmetries in the outer regions. 

\begin{figure}
\plotone{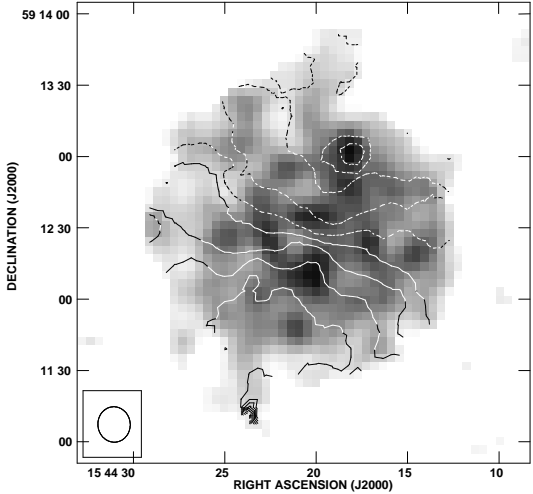}
\caption{The velocity field of SBS 1543+593 showing the velocity contours
overlaid on a greyscale map of the \hi. The contours are separated by 10 
\kms\ and main contours range from 2838 to 2898 \kms\ with the smallest 
velocities on the southern side of the galaxy.}
\label{fig:sbsvelmap}
\end{figure}

\begin{figure}
\plotone{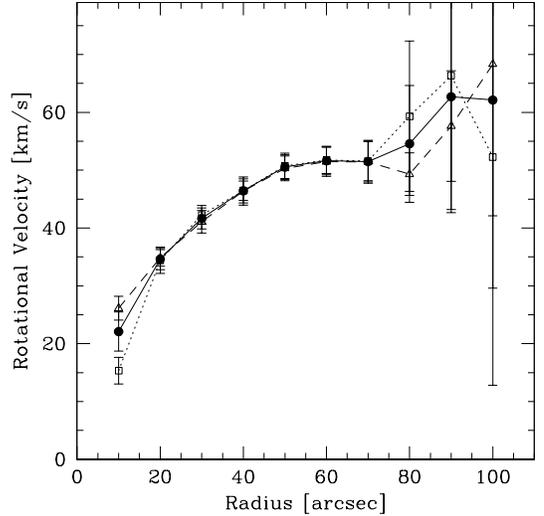}
\caption{The rotation curve of SBS1543+593. Fits on the northern (receding, open
squares) and southern (approaching, open triangles) halves of the galaxy were 
performed separately and are plotted as the dotted and dashed lines
respectively. The filled circles and solid line correspond to the fit for the
entire galaxy.}
\label{fig:rotcur}
\end{figure}

Figure \ref{fig:sbsvelmap} shows the \hi\ column density distribution 
in greyscale overlaid with the \hi\ velocity contours. The figure shows a
regular velocity gradient as would be expected from the rotation of an
inclined spiral galaxy. However, note that the velocity of region 1 is set off
by closed contours, not exactly following the rotation curve of the galaxy. 
Nevertheless, this is not a large deviation from the rotation velocity as it is
not visible in the rotation curve of the Northern side of the galaxy as shown
in Figure \ref{fig:rotcur}. The velocity field as produced by AIPS was ported to 
GIPSY \citep{vanderhulst1992} and was used to determine the rotation curve and 
orientation (inclination and position angle) of the galaxy. We used the task
ROTCUR to perform a least squares fit to the radial velocity field solving for
the following parameters: the position of the dynamical center, and the
systemic velocity. Fits were made using the conventional approach -- reasonable 
starting values were provided for the galaxy center and its systemic velocity 
and were kept fixed while fits were made in 10\arcs\ wide annuli to solve for the
inclination, position angle and rotation curve. These resulting values were then 
kept fixed and ROTCUR was left to determine improved solutions for the position 
of the dynamical center and the systemic velocity. This was repeated until
the process converged, resulting in the position of the kinematic
center (in excellent agreement with the position of the optical
nucleus), the systemic velocity (v$_{sys}$=2867.4 km/s), inclination ($i =
40$\deg), and position angle (PA = 344\deg). These parameters are consistent
with the values derived by \citet{chengalur2002} -- v$_{sys}$ = 2870 \kms, $i =
50$\deg, and PA = 344\deg. Several runs were made, allowing
one or more parameters to vary, investigating for example if there are
indications for a warp. There was no compelling case, though, for
either the inclination or position angle to vary with radius. The
resulting rotation curve is plotted in Figure \ref{fig:rotcur}. In
order to check for asymmetries we performed a fit on the northern and
southern halves of the galaxy separately and these results are plotted
as well (dotted and dashed line). We derive slightly higher rotation
velocities at each point than \citet{chengalur2002} -- we find an inclination
corrected rotation speed
at 60\arcs\ of v$_{rot}$ = 52 \kms\ corresponding to a dynamical mass of
M$_{dyn}$(60) = 7$\times 10^9$ M\solar\ as compared with their value of
M$_{dyn}$(60) = 5$\times 10^9$ 
M\solar. From our measurements the rotation curve appears to be flat or slightly
rising out to at least 100\arcs\ indicating M$_{dyn}$(100) \gapprox 1$\times 10^{10}$ 
M\solar.

Figure \ref{fig:sbsmap} shows that the \hi\ disk of SBS 1543+593 is much larger 
than the optical extent. The \hi\ to optical size ratios, D$_{HI}$/D$_{25}^B = 15$ 
and D$_{HI}$/D$_{6.4h} = 2$ indicate that the galaxy has an extended \hi\ disk
relative to its stellar light. 
Spiral galaxies typically have \hi\ disks that are extended relative to the 
stellar disks with values of 1.7 for both of these ratios \citep{broeils1997}. 
Dwarf galaxies tend 
to have even more extended \hi\ disks with values of 3.3 and 1.8 for these ratios 
respectively \citep{swaters2002} making this a fairly normal gas-rich 
low surface brightness galaxy once the extremely low surface brightness has been
accounted for. In
addition, SBS 1543+593 is unusually faint for a galaxy with this mass of HI:
M$_{HI}$/L$_B = 4 \pm\ 1$ while the average value is only 1.5 for late-type dwarf 
galaxies \citep{swaters2002}. SBS 1543+593 and its companions  also fall along the 
HI size -- HI mass relationship for galaxies from \citet{rosenberg2003}. SBS
1543+593 is slightly larger than most objects along that relation, but not by a 
significant margin.

With an R-band measurement and a (B-R) color with a large uncertainty (0.2 
$\pm$ 0.3) for SBS 1543+593 the stellar mass of the galaxy can not be determined, 
but the Padova 1994 models of \citet{bruzual2003} can be used to place limits 
on the stellar population. These limits are based on the assumption of a Salpeter 
initial mass function, a metallicity of 0.2Z$_{\odot}$ (the closest model 
metallicity to the present day abundance of 0.3Z$_{\odot}$, \citealp{schulte2005}), 
and a single stellar population. The upper limit to the stellar mass adding the
assumption that the stellar population is 13.7 Gyr old is 8.1$\times 10^8$ 
M\solar, but the model galaxy color, (B-R) = 0.82, is too red compared with the 
measured color indicating that the stellar population is not maximally old. 
Alternatively, the lower limit to the stellar
mass is 4.2$\times 10^6$ M\solar\ assuming an age for the stellar population of
3.5 Myr for which (B-R) = -0.08 from the models which is consistent with the
galaxy color within the errors. Assuming the measured (B-R) color to identify
a ``best guess" model gives an age for the stellar population of 2.2 
$\times 10^8$ years and a stellar mass of 4.5 $\times 10^7$ M\solar. For 
this best-guess stellar mass the ratio of gas mass to stellar mass is 27 -- the
baryonic content of SBS 1543+593 is dominated by its gas. The optical light 
in the galaxy is dominated by stars that have formed recently. However, 
because it is impossible to rule out the existence of an underlying old stellar 
population with the available information, this may be a young galaxy or it may
be one that has been extremely inefficient in turning its gas into stars.

Systems like SBS 1543+593 with extended disks and high values of M$_{HI}$/L$_B$
are rare in the universe. Galaxies with M$_{HI}$/L$_B > 1$ are only
expected to contribute 5\% of the \hi\ cross-section \citep{ryan-weber2003} in the
local universe. While the extended disk of SBS 1543+593 could make it more 
readily detectable as a DLA than other galaxies, that did not play a role in the
detection of this systems since the QSO is very close to the galaxy's center. 

Overall the \hi\ disk of SBS 1543+593 is does not show evidence for strong star
formation -- it is optically low 
surface brightness even in the regions of high \hi\ column density. The second
contour in Figure \ref{fig:sbsmap} is only a factor of 2 below the upper end of 
the hydrogen mass surface density threshold for star formation which includes
the molecular gas component \citep{kennicutt1989}. All of the
high column density regions in this system show stellar emission, but for
several regions, the emission is extremely faint.
\citet{schulte2005} estimate a star-formation rate of only 0.006 $h_{70}^{-2}$
M\solar\ yr$^{-1}$ over the whole galaxy. In addition, \citet{bowen2005} 
have used the column density measurement at the position of the QSO to place a 
limit on the star-formation rate surface density of 7$_{-4}^{+9}$ M\solar\ 
yr$^{-1}$ kpc$^{-2}$. The absorption line measurements using the STIS ultraviolet 
spectrograph on the Hubble Space Telescope from \citet{bowen2005} indicate that the 
line profiles in this central region are simple, also showing no evidence for more 
than a single absorption line component in the low ionization species (at 20 km/s 
resolution). If there was an outflow at a large distance above the plane there
would be a chance of intersecting it along the line of sight to the QSO (the
path is slightly longer than it would be for a face-on system). As with the HI
emission, the absorption lines do not show evidence for starburst induced outflows 
in the inner region of this galaxy. All in all, there is only a low level of
star formation in this system. 

\subsection{Optical and \hi\ Properties of SBS 1543+593's Neighbors}

\begin{figure}[ht]
\plotone{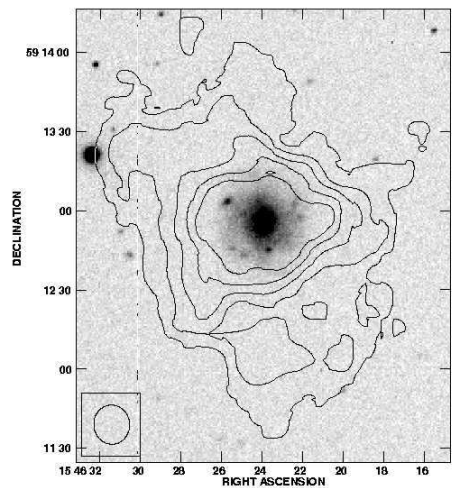}
\caption{The $V$-band image of MCG+10-22-038 overlaid 
with \hi\ contours from the VLA map. The lowest level contour represents 
2$\times 10^{20}$ cm$^{-2}$ column density where Lyman-$\alpha$ absorbers are 
classified as DLA systems. The rest of the contours are at 6, 10, 14, and 18 
$\times 10^{20}$ cm$^{-2}$. The circle in the lower left corner shows the beam
size for these observations which is 14.9\arcs$\times$13.6\arcs.}
\label{fig:mcgmap}
\end{figure}

\begin{figure}[ht]
\plotone{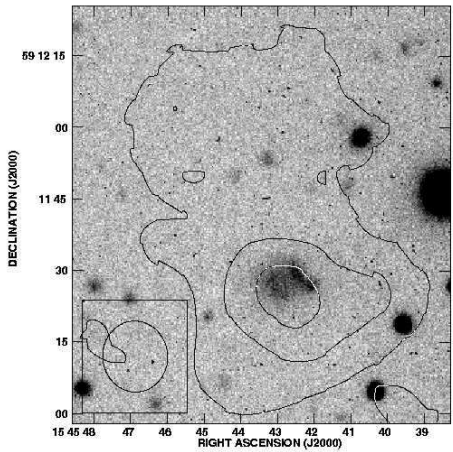}
\caption{The \hi\ contours from the VLA map of Dwarf 1 
(RBTB154542.8+591132)
overlaid on the $R$-band image. The lowest level contour represents 2$\times 
10^{20}$ cm$^{-2}$ column density where absorbers are classified as DLA systems. 
The other contours are at 6 and 10 $\times 10^{20}$ cm$^{-2}$. The circle in the 
lower left corner shows the beam size for these observations which is 
14.9\arcs$\times$13.6\arcs.}
\label{fig:dwarfmap1}
\end{figure}

\begin{figure}[ht]
\plotone{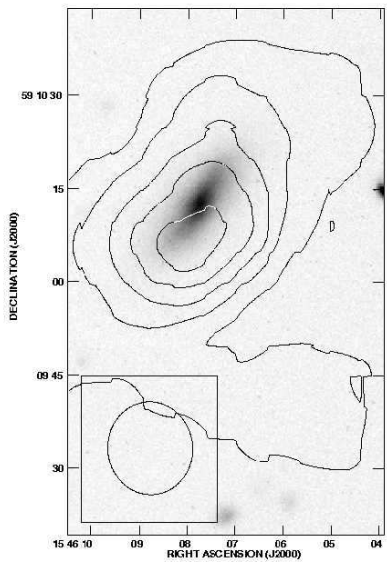}
\caption{The \hi\ contours from the VLA map of Dwarf 2 (RBTB154607.7+591013)
overlaid on the $V$-band image. The lowest level contour represents 2$\times 
10^{20}$ cm$^{-2}$ column density where absorbers are classified as DLA systems. 
The other contours are at 6, 10, and 14 $\times 10^{20}$ cm$^{-2}$. The $R$-band 
image is shown in the right-hand panel. The circle in the lower left corner shows 
the beam size for these observations which is 14.9\arcs$\times$13.6\arcs.}
\label{fig:dwarfmap2}
\end{figure}

MCG+10-22-038, Dwarf 1, and Dwarf 2 were all detected in
the VLA observations of SBS 1543+593 because of their \hi\ content.
The left-hand panel of Figures \ref{fig:mcgmap}, \ref{fig:dwarfmap1}, and 
\ref{fig:dwarfmap2} show the VLA maps of these three systems overlaid on the
optical images. The \hi\ in these galaxies is less extended relative to their
optical scale lengths than SBS 1543+593 yet all of the galaxies are extremely
gas-rich systems. 

Figure \ref{fig:mcgmap} shows that MCG+10-22-038 is a face-on low 
luminosity but moderate surface brightness disk with a few bright HII regions. 
The galaxy is less gas-rich than 
its neighbors with M$_{HI}$/L$_B = 0.5$, but this value is consistent with the
average value for spiral galaxies \citep{broeils1997}. 

Dwarf 1 is the most unusual system in this small group. The galaxy is 
faint (M$_V = -13.8$), low surface brightness ($\mu_0$(B) $= 24.2$),
blue ($V-R = 0.3$), and is a factor of 3 more gas-rich than any of the 
late-type dwarf galaxies in the sample of \citet{swaters2002} with M$_{HI}$/L$_B
= 5$. The galaxy is faint and irregular with two higher surface brightness bands 
superposed on the faint disk (\ref{fig:dwarfmap1}). Another irregularity in the 
system is that the \hi\ 
extends to the north of the optically visible part of the galaxy. This galaxy 
has remained low luminosity and low surface brightness despite a 
reservoir of star-forming material and the presence of 2 close neighbors.

Dwarf 2 is an edge-on dwarf spiral as can be seen in Figure \ref{fig:dwarfmap2}. 
The galaxy is low luminosity (M$_R = -16.4$) with a redder $V-R$ color 
than the other galaxies in the vicinity either due to an older stellar 
population, extinction exacerbated by its orientation, or both. This
galaxy has a lower total \hi\ mass and a lower peak \hi\ column density than the
other systems but an M$_{HI}$/L$_B$ that is higher than the value for
MCG+10-22-038, the other dwarf spiral in the group. In Dwarf 2, the gas follows 
the optical contours showing only a slight irregularity in the lowest
signal-to-noise contours which are not highly reliable because of the noise
which increases with distance from the center of the map.  

The neighbors to SBS 1543+593 are, overall, gas-rich systems -- MCG+10-22-038 is
slightly below average for late-type dwarf galaxies, Dwarf 2 is approximately 
average, and Dwarf 1 has 
M$_{HI}$/L$_B = 5$, much more gas-rich than an average late-type dwarf galaxy
for which M$_{HI}$/L$_B = 1.5$ \citep{swaters2002}. All together, these systems
have an \hi\ cross-section to damped Lyman-$\alpha$ absorption that is $\sim$70\% 
of the cross-section of SBS 1543+593. Galaxies with \hi\ masses of 
$< 10^9$ M\solar\ make up about 42\% of the expected \hi\ cross-section at $z = 0$ 
\citep{rosenberg2003} so field dwarfs and/or small galaxy groups like this one
are fairly common and are important contributors to the DLA
population. At high redshift where young, gas-rich groups were more common, these
kinds of systems may be even more significant contributors to the DLA cross-section.

\subsection{\hi\ Environment}

The small scale environment (of order 200 kpc) of SBS 1543+593 includes 3 low 
luminosity galaxies that have been identified in our VLA \hi\ observations. These 
galaxies -- MCG+10-22-038, Dwarf 1, and Dwarf 2 -- make up a tight group with
projected distances of 183 kpc, 123 kpc, 
161 kpc from SBS 1543+593 respectively. For the small detection volume covered 
by our VLA map, a conservative estimate of the average galaxy density 
would predict 8.6$\times 10^{-3}$ galaxies down to log(M$_{HI}$/\msolar) $=
8.07$, the detection limit of these observations at the distance of the detected 
galaxies, 
in the field using the \hi\ mass function from \citet{rosenberg2002}. Because this 
region is not an unbiased position -- since it was centered around a known galaxy 
and galaxies tend to cluster -- a higher than average galaxy density should not 
be surprising. However, the detection of 3 galaxies in the immediate vicinity 
of SBS 1543+593 indicates a significant overdensity with respect to the field. 
The high density is also significant given that LSB galaxies 
tend to have fewer near neighbors than their higher surface brightness
counterparts \citep{taylor1997}. While the galaxy overdensity in this region is 
statistically significant, there is a large variation in the number of nearest 
neighbors to low surface brightness galaxies ($\mu_0$(B) $\ge$ 23 mag arcsec$^{-2}$) 
on scales less than 0.5 Mpc 
\citep{bothun1993}, and the expected number of low luminosity gas-rich neighbors
is entirely unknown. 

Finding DLAs in regions that contain several low luminosity galaxies is not
unusual. This then implies the presence of several low luminosity systems makes
it much more difficult at high redshift to unequivocally identify which object
is responsible for the
absorption. \citet{steidel1997} find a DLA at $z = 0.656$ for which the associated 
galaxy is unidentified, but there are several other galaxies in the vicinity 
including a
dwarf with L$_K = 0.07$L$_K^*$ which is more than 140 kpc from the line of
sight. This system is at too high a redshift to be studied at 21cm so the 
gaseous properties of the system are unknown, but otherwise the environment of
this DLA appears to be similar to that of SBS 1543+593. As we look deeper both 
in the optical and at 21 cm, we 
may find that it is not uncommon for DLAs to be associated with young gas-rich 
galaxy groups. In fact, the combination of rings, tidal features and gas-rich
neighbors could contribute to the multiple
kinematic components often seen in high resolution DLA observations
\citep{wolfe2000}. The presence of a significant amount of gas surrounding
galaxy groups is also consistent with the determination that most of the low
column density \hi\ seen in absorption in the Lyman-$\alpha$ forest is also
associated with galaxy groups \citep{ryanweber2006,morris2006}.

\begin{figure}[ht]
\plottwo{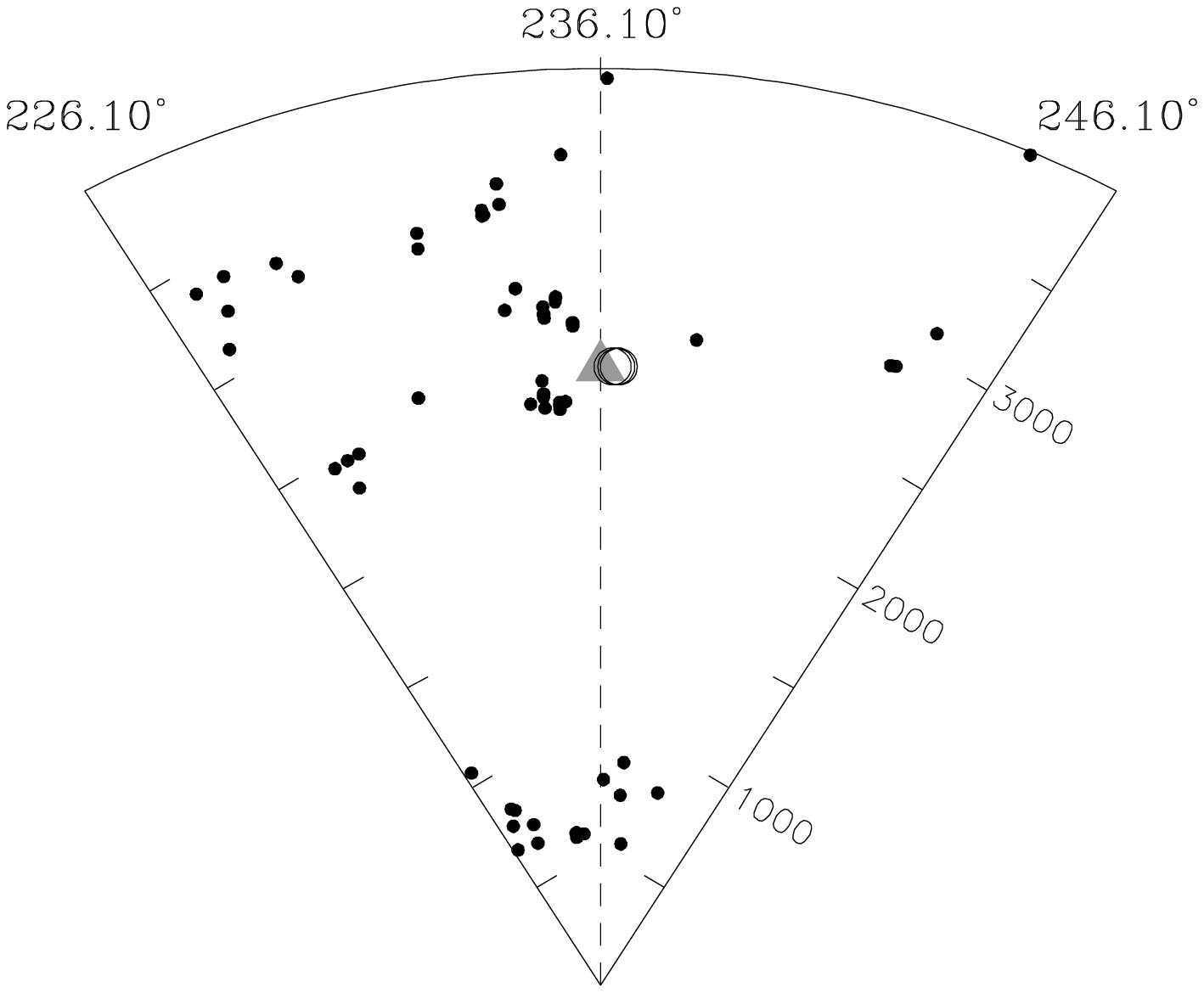}{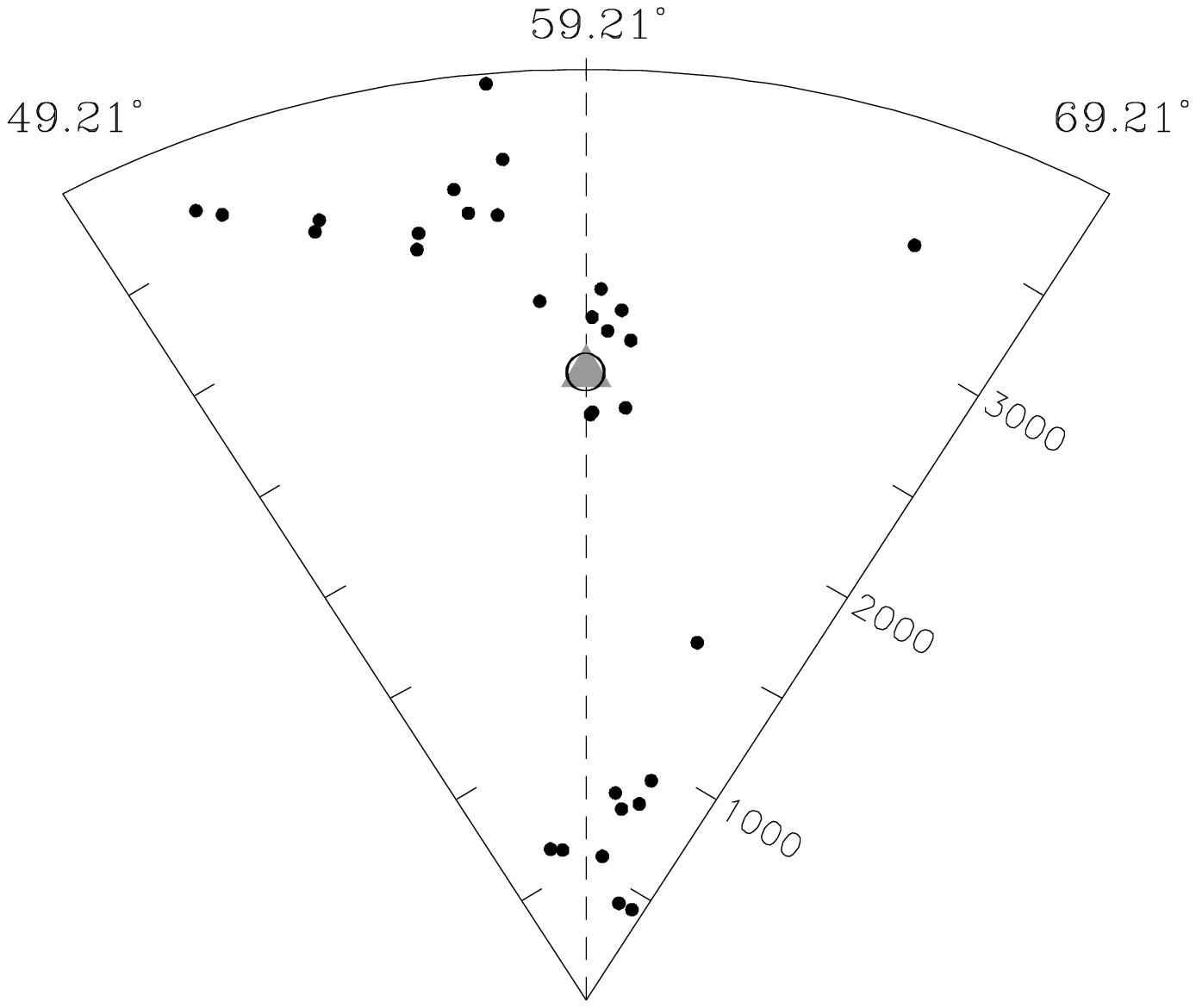}
\caption{The right ascension (left-hand plot) and declination (right hand plot)
cone diagrams for this region. In the left-hand plot, right ascension is along
the top and velocity is along the diagonal. In the right-hand plot, declination 
is along the top and velocity is along the diagonal. The small filled circles
are galaxies from the ZCAT 
catalog, which includes data from the CfA survey as well as from several other 
galaxy surveys, within $\pm10$\deg\ of SBS 1543+593. The grey triangle shows 
the position of SBS 1543+593 while the open circles show the positions of the 
newly identified galaxies (the points fall nearly on top of one another). The
dashed line indicates the line of sight to QSO HS 1543+5921 in which the damped
Lyman-$\alpha$ absorption was identified. Note that on these scales the galaxies
in this group are very close together and can not easily be distinguished.}
\label{fig:cones}
\end{figure} 

On large scales, low surface brightness galaxies ($\mu_0$(B) $\ge$ 23 mag 
arcsec$^{-2}$) are found to follow the
overall large scale structure, but are slightly less clustered than their
higher surface brightness counterparts \citep{mo1994}.
Figure \ref{fig:cones} shows the distribution of galaxies from the ZCAT catalog
\footnote[1]{http://cfa-www.harvard.edu/$\sim$huchra/zcat/} \citep{falco1999}, 
which includes data from the CfA survey as well as from several other galaxy 
surveys, in the region surrounding 
SBS 1543+593 (small black filled circles). The grey triangle shows the position
of SBS 1543+593 while the open circles show the positions of the newly
identified galaxies (the points fall nearly on top of one another). The dashed
line indicates the line of sight to the QSO HS 1543+5921. The plot shows that 
these new galaxies form a tight group (the individual galaxies can not be
distinguished on this scale, see Figure 1 for the relative positions within the
group) which in projection on the sky measures
$\sim$185 kpc from SBS 1543+593, significantly closer than any of the previously 
known higher surface brightness systems. All of these systems sit within a
filament of galaxies that stretches off to the southwest of SBS 1543+593.

Despite living in a dense galactic environment, these galaxies are low
luminosity and low surface brightness indicating that only a small fraction of
the available gas  has been turned into stars. That lack of a significant amount 
of star-formation indicates that either the group is very young and has not had 
time to interact with the other galaxies or the timescales for interaction are 
long. The timescale that we can examine for this system is the crossing time 
derived 
from the velocity dispersion of the system. With a velocity dispersion of 132 
\kms\ and a size of 183 kpc, the crossing time is 1.4$\times 10^9$ yr. It might 
be expected that the tight sub-group of three neighbors -- MCG+10-22-038,
Dwarf 1, and Dwarf 2 -- is a more likely location for strong interactions, but 
the velocity dispersion is only 19 \kms\ making the crossing time 
3.3$\times10^9$ yr over the region's 63 kpc size. In either case the timescales 
are a few Gyrs indicating that these may not be extremely young objects, but if
this group did form at a redshift significantly less than one as this implies,
some low mass, gas-rich galaxy groups may have been assembled at fairly late
times.

\section{Discussion}

We have used the VLA in C-array to map the \hi\ distribution in and around one
of the
nearest DLAs beyond the local group, SBS 1543+593. We identify 2 previously
uncataloged galaxies and find a redshift for MCG+10-22-038. All of the galaxies
are found to be low luminosity, gas-rich systems with M$_{HI}$/L$_B$ 
ranging between 0.5 and 5. 

The low surface brightness of SBS 1543+593, despite a large reservoir of fuel
for star formation, indicates that this galaxy is young or has not undergone a
strong interaction in its recent past. However, SBS 1543+593 is not an isolated
galaxy -- on both small and large scales this is a fairly high density region of
the universe yet the interaction timescale with its neighbors is
moderate (1.4 Gyr). 

Gas-rich, but still quiescent galaxy groups in the local universe are very hard
to find because they are low luminosity systems, but they provide important 
information for understanding gas-accretion
and star-formation processes in these systems. The accretion of cold gas onto 
dwarf galaxies at late times \citep{mo2005, keres2005} is predicted to be an 
important mechanism for building up gas in dwarf galaxies and theoretically it
can occur without triggering a major star-formation episode. The fueling of this 
accretion process by gas in the intergalactic
medium surrounding this group is also consistent with the observations that most
of the Lyman-$\alpha$ absorbers, which probe the intergalactic medium, are
associated with galaxy groups \citep{ryanweber2006}. 

If gas-rich galaxies were more likely to form at high redshift when the density
of gas in the intergalactic medium was higher, then SBS 1543+593 may provide 
important information about the environment in which many of these DLAs may 
reside. These data also provide a warning about how easy it
is to overlook dwarf galaxies even in the local universe. This is a nearby
DLA that was previously thought to be isolated. In the high redshift
universe it becomes extremely difficult to detect galaxies like these which have
luminosities that are a small fraction of L$_*$ -- MCG+10-22-038, the highest
luminosity system in the group, is only $\sim$0.13L$_{V_*}$ 
(adopting the value of M$_{V_*}$ = -19.91 from \citealp{delapparent2003}).

DLAs cover a wide range of morphologies and surface brightnesses but it has
become clear that low surface brightness systems are an important contributor to
the population (e.g., \citealp*{rosenberg2003,bowen2001a,bouche2001,steidel1994, 
lanzetta1997,lebrun1997, pettini2000, turnshek2001, colbert2002, kulkarni2005}). 
Often low surface brightness galaxies
are thought to be isolated systems because they are less strongly clustered on
large scales. However, on small scales the clustering around these
galaxies is less well determined especially when considering the role of
gas-rich galaxies that might have been missed in optically-selected samples.
Learning more about the local environment of gas rich galaxies and constraining 
the models of gas accretion on to galaxies as well as to understand
their local gaseous environments on small scales requires a better understanding
of the distribution of low mass, gas-rich galaxies in the local universe. The
next generation of \hi\ surveys for galaxies will provide both the sensitivity
and the resolution needed to address these questions \citep{giovanelli2005a} in
much greater detail. The goal will be to use this improved understanding of the
gas distribution surrounding local galaxies to help interpret the observations
of more distant gas-rich objects identified in DLA studies or by other means.

\acknowledgements
We thank John Salzer for providing us with the KPNO calibration. We would also
like to thank Regina Schulte-Ladbeck and Brigitte K\"onig for providing us with
their GMOS image for comparison with the \hi\ data prior to its official release 
from the Gemini archive. 
JLR acknowledges support from NSF grant AST-0302049. DVB acknowledges support from 
NASA grant HST-GO-09784 from the Space Telescope Science Institute, which is operated
by the Association of Universities for Research in Astronomy, Inc., under NASA 
contract NAS5-26555; and from NASA grant NNG05GE26G. TMT acknowledges support
from NASA grant NNG-04GG73G. Some of this work was based on observations obtained 
with the Apache Point Observatory 3.5-meter telescope, which is owned and operated 
by the Astrophysical Research Consortium.This paper made use of the Second Palomar 
Observatory Sky Survey (POSS-II) which was made by the California Institute of 
Technology with funds from the National Science Foundation, the National Geographic 
Society, the Sloan Foundation, the Samuel Oschin Foundation, and the Eastman
Kodak Corporation.

%\bibliographystyle{astron}
%\bibliography{references}

\clearpage
\begin{landscape}
\begin{deluxetable*}{lllccrcc}
\tabletypesize{\small}
%\rotate
\tablewidth{7.4in}
\tablecaption{\hi\ Data}
\tablehead{
\colhead{Name} & \colhead{$\alpha$} & \colhead{$\delta$} & \colhead{S} &
\colhead{Vel.} & \colhead{$\Delta$V$_{50}$ \tablenotemark{a}} & \colhead{M$_{HI}$} &
\colhead{Sep. \tablenotemark{b}} \\
\colhead{} & \colhead{(J2000)} & \colhead{(J2000)} & \colhead {(Jy)} & \colhead{(\kms)} &
\colhead{(\kms)} & \colhead{(\msolar)} & \colhead{(kpc)}}
\startdata
SBS 1543+593 & 15:44:20.3 \tablenotemark{c} & 59:12:24 & 0.63 & 2864 & 75 & 1.2$\times 10^9$ & \nodata \\
MCG+10-22-038 & 15:46:24.6 & 59:12:54 & 0.22 & 2685 & 98 & 1.1$\times 10^9$ & 184 \\
RBTB J154542.8+591132(Dwarf 1) & 15:45:42.8 & 59:11:32 & 0.34 & 2710 & 42 & 1.8$\times 10^8$ & 126 \\
RBTB J154607.7+591013(Dwarf 2) & 15:46:07.7 & 59:10:13 & 0.13 & 2690 & 122 & 2.0$\times 10^8$ & 166 \\
\enddata
\tablenotetext{a}{The full velocity width of the line measured at 50\% of the peak flux.}
\tablenotetext{b}{The impact parameter between SBS 1543+593 and the galaxy}
\tablenotetext{c}{Coordinates from \citet{schulte2004}}
\end{deluxetable*}

\begin{deluxetable*}{lrcccccccrrrrr}
\tabletypesize{\tiny}
%\rotate
\tablewidth{9in}
\tablecaption{Optical Data}
\tablehead{
\colhead{Name} & \colhead{Radius[V] \tablenotemark{a}} & \colhead{V} & \colhead{R} & \colhead{M$_V$} & 
\colhead{M$_R$} & \colhead{$\mu_0$(V) \tablenotemark{b}} & \colhead{$\mu_0$(B)} & \colhead{D$_{25}^B$} & 
\colhead{D$_{h}$ \tablenotemark{c}} & \colhead{V-R} & \colhead{M$_{HI}$/L$_B$} & 
\colhead{D$_{HI}$/D$_{25}^B$ \tablenotemark{d}} & \colhead{D$_{HI}$/D$_{6.4h}$
\tablenotemark{e}}\\
\colhead{} & \colhead{(\arcs)} & \colhead{} & \colhead{} & \colhead{} & \colhead{} &
\colhead{(mag arcsec$^{-2}$)} & \colhead{(mag arcsec$^{-2}$)} & \colhead(\arcs) & 
\colhead{(\arcs)} & \colhead{} & \colhead{} & \colhead{} & \colhead{}}
\startdata
SBS 1543+593 \tablenotemark{f}    & 33.3                 & \nodata  & 16.3   & \nodata & -15.9  & 22.6  & 23.2 & 9.6 & 14.6   & \nodata & 4   & 15  & 2\\
MCG+10-22-038                  & 27.5                & 15.2   & \nodata  & -17.7  & \nodata & 21.4  & 21.7 & 34.6 & 15.6 & \nodata & 0.5 & 4   & 2\\
RBTB J154542.8+591132(Dwarf 1) & 9.4                 & 19.1   & 18.8    & -13.8 & -14.0  & 23.8  & 24.2 & 11.0 &  7.5 & 0.3    & 5   & 10 & 2\\
RBTB J154607.7+591013(Dwarf 2) & 22.0 \tablenotemark{g}  & 17.2   & 16.5   & -15.7  & -16.4  & 21.3  & 22.2 & 24.0 &  7.5 & 0.7   & 0.9 & 3   & 2\\
\enddata
\tablenotetext{a}{Galaxy diameter in V-band except for SBS 1543+593 for which it is
the R-band size}
\tablenotetext{b}{V-band central surface brightness}
\tablenotetext{c}{The size at 6.4 times the optical scale-length}
\tablenotetext{d}{The ratio of H I size at 1 M\solar\ pc$^{-2}$ to the size at 25
mag arcsec$^{-2}$}
\tablenotetext{e}{The ratio of H I size at 1 M\solar\ pc$^{-2}$ to the size at
6.4 times the optical scale length}
\tablenotetext{f}{Data for SBS 1543+593 are from \citet{bowen2001a}}
\tablenotetext{g}{This value is the semi-major axis of the aperture which had an
ellipticity of 0.52 and a position angle of -29.3.}
\end{deluxetable*}
\clearpage
\end{landscape}

\end{document}